\documentclass[11pt,a4paper]{article}
\usepackage{jcappub}
\usepackage{hyperref}
\usepackage{subfig}
\hypersetup{colorlinks = true, citecolor = blue}
\usepackage{cancel}

\newcommand{\bvec}[1]{\vec{#1}}

\newcommand{\al}[1]{\begin{align} #1 \end{align}}
\newcommand{\eq}[1]{\begin{equation} #1 \end{equation}}

\begin{document}
%
%
\title{On decoherence of cosmological perturbations \\ and stochastic inflation
       }
\author{Jan Weenink}
\emailAdd{j.g.weenink@uu.nl}
\author{and Tomislav Prokopec}
\emailAdd{t.prokopec@uu.nl}
\affiliation{Institute for Theoretical Physics and Spinoza Institute,\\
Utrecht University, Leuvenlaan 4, 3585 CE Utrecht, The Netherlands}
\abstract{
 By making a suitable generalization of the Starobinsky  stochastic
inflation, we propose a classical phase space formulation
of stochastic inflation which may be used for a quantitative study
of decoherence of cosmological perturbations
during inflation. The precise knowledge of how much cosmological perturbations
have decohered is essential to the understanding of acoustic
oscillations of cosmological microwave background (CMB) photons.
In order to show how the method works, we provide the relevant equations for
a self-interacting inflaton field. For pedagogical reasons and to provide a link
to the field theoretical case, we consider the quantum stochastic harmonic oscillator.
}
\arxivnumber{1108.3994}
\maketitle
\section{Introduction}

Decoherence~\cite{Zeh:1970,Joos:1984uk,Zurek:2003zz} describes how a quantum
system evolves into a state which most closely resembles a classical state
through an interaction with the environment.
As information about the system is lost to the environment as perceived by some observer,
the entropy of the system increases.
Recently quantitative studies have been done on decoherence for interacting
quantum field theories, see
Refs.~\cite{Campo:2008ju,Campo:2008ij,Giraud:2009tn,Koksma:2009wa,Koksma:2010zi,Koksma:2011dy,Gautier:2011fx}.
These studies rely on the fact that neglecting observationally inaccessible, non-Gaussian
correlators will give rise to an increase in Gaussian entropy of the state,
which is then taken as a quantitative measure of decoherence.
A useful application of this approach is to quantitatively study
the decoherence of cosmological perturbations, \textit{i.e.}
cosmological decoherence.
During inflation quantum fluctuations of the inflaton field -- and thus
scalar cosmological perturbations -- generically
decohere because of the interactions generated by the inflationary potential
as well as because of the interactions between cosmological perturbations
and other fields present during inflation, and as a result become
more and more
classical~\cite{Brandenberger:1992sr}.\footnote{In Ref.~\cite{Polarski:1995jg}
it is stated that even in the absence of interactions
the quantum fluctuations become indistinguishable from classical fluctuations
during inflation. The authors call this "decoherence without decoherence".
Technically, the state becomes extremely squeezed,
but the phase space area remains constant (pure state).
Our notion of decoherence differs, in that we
associate decoherence with growth of the phase space
area and consequently the entropy of the state.}
The ultimate goal of this project is to find out
what the effect of cosmological decoherence on the cosmological
microwave background (CMB) spectrum is for specific inflationary
models, and thus to make a connection between
inflationary models and late time cosmological observables.

 In this paper we focus on cosmological decoherence in the framework
of stochastic
inflation~\cite{Starobinsky:1986bb,Starobinsky:1994bd,Tsamis:2005hd}.
In stochastic inflation the system is represented by the super-Hubble (IR)
modes of the inflaton field, as these modes are amplified during inflation
and they are the ones that ultimately induce temperature fluctuations
in the CMB. The sub-Hubble (UV) modes form the environment and act as
a (white) noise for the long-wavelength modes.

Central in the stochastic inflation scenario is the Starobinsky equation
\eq{
\dot{\phi}(t,\bvec{x})=-\frac{1}{3H}V'\left(\phi(t,\bvec{x})\right)+F(t,\bvec{x}).\label{Starobinskyequation}
}
The field $\phi$ is the coarse-grained part of the complete field,
\textit{i.e.} it contains only the long wavelength modes with $k\ll Ha$.
$H=\dot{a}/a$ ($a$ is the scale factor) is the Hubble parameter
in inflation and $V(\phi)$ is a potential, {\it e.g.}
$V(\phi)=\lambda \phi^4/4!$ for a self-interacting scalar field in
a chaotic inflationary scenario. The noise term $F$ originates from
the short wavelength modes which keep crossing the Hubble radius
due to the accelerated cosmic expansion. Interactions between
UV modes are neglected, such that
these modes are uncorrelated over time, thus the noise is white,
\eq{
\langle F(t,\bvec{x}) F(t',\bvec{x}) \rangle = \frac{H^3}{4\pi^2}\delta(t-t'),
}
where $\langle \cdot \rangle$ denotes a (statistical) average.
Although the coarse grained field $\phi$ is a quantum field with
creation and annihilation operators, in inflation it nevertheless commutes
with its first time derivative ($\propto$ the canonical momentum).
In this sense the field $\phi$ may be called
a classical stochastic field~\cite{Tsamis:2005hd}.
The Starobinsky equation is therefore a classical Langevin equation
for the stochastic field $\phi$,
in contrast to the quantum field equation for the complete quantum scalar
field. From Eq.~\eqref{Starobinskyequation} a Fokker-Planck equation for
the probability distribution of $\phi$ can be derived,
which is then used to calculate correlators for $\phi$.

Perhaps the greatest success of stochastic inflation is that it recovers
the leading infrared logarithms (\textit{i.e.} $\ln^n(a)$ terms) of
the field correlators at each order in perturbation
theory~\cite{Starobinsky:1986bb,Tsamis:2005hd,Miao:2006pn,Prokopec:2006ue,Prokopec:2007ak,Prokopec:2008gw}.
Originally it was believed that the time dependent separation between short
and long wavelength modes would also lead to cosmological decoherence, even
in a free field theory (see \textit{e.g.}
Refs.~\cite{Hosoya:1988yz,Nambu:1991vs}).
However, Habib~\cite{Habib:1992ci} showed that the free stochastic equations
do not lead to any decoherence.

Although the Starobinsky equation~\eqref{Starobinskyequation}
is excellent for describing the leading order behavior of the coincident
field correlators, it fails when one is interested in the phase space of a state.
Technically the Starobinsky equation cannot be derived from an action,
thus no canonical momentum and Hamiltonian can be constructed.
Moreover from the Starobinsky equation~\eqref{Starobinskyequation} no off-coincident two-point
functions can be derived. This is especially troublesome for our approach
to decoherence from the viewpoint of increased entropy due to incomplete
knowledge of observationally inaccessible higher
correlators~\cite{Campo:2008ju,Campo:2008ij,Giraud:2009tn,Koksma:2009wa,Koksma:2010zi,Koksma:2011dy}.
The entropy of the system is in this case described by
the Gaussian von Neumann entropy~\cite{Koksma:2010zi,Campo:2008ju,Campo:2008ij}
\eq{
S_{\text{g}}=\int \frac{d^{3}\bvec{k}}{(2\pi)^{3}} \left[\frac{\Delta_k+1}{2}\ln\left(\frac{\Delta_k+1}{2}\right)
-\frac{\Delta_k-1}{2}\ln\left(\frac{\Delta_k-1}{2}\right)\right],
\label{EntropyvonNeumann}
}
where $\Delta_k$ is the phase space area ($\hbar=1$) for a Gaussian
state centered at the origin,
\eq{
\Delta_k^2=4\left[\langle |\phi(t,\vec k)|^2 \rangle
                  \langle |\pi_{\phi}(t,\vec k)|^2 \rangle
        -\langle \frac12(\phi(t,\vec k)\pi_{\phi}^{\star}(t,\vec k)+\pi_{\phi}(t,\vec k)\phi^{\star}(t,\vec k)) \rangle^2
            \right]
\,,
\label{phasespacearea}
}
where $\phi(t,\vec k)$ is a Fourier mode of the field $\phi(t,\vec x)$
that describes the system, and $\pi(t,\vec k)$ is its conjugate momentum.
The phase space area is equal to 1 for a pure state, which consequently
has zero entropy. For a mixed state the phase space area grows
and entropy increases. Mixing occurs in general for an interacting
quantum field theory, such as a self-interacting quartic potential
for the inflaton field, used by Starobinsky and Yokoyama~\cite{Starobinsky:1994bd}.
However, because the canonical momentum is absent in the Starobinsky
equation~\eqref{Starobinskyequation}, the system will be in
a 'pencil state' with zero phase space area. The entropy for such a state is not defined.
Obviously we have to improve the Starobinsky equation in order to describe
the growth of the phase space area and entropy.


The solution is a full Hamiltonian (phase space) formulation of
stochastic inflation, which was first (for a different purpose) used
by Habib~\cite{Habib:1992ci} for the case of free quantum fields.
The idea is to introduce coarse-graining for both the scalar field
and its conjugate momentum. As a result there will
be two noise terms in the Hamilton equations for the IR field and its momentum (see also Ref. \cite{Tolley:2008na} for a more recent discussion). 
This is very much like a quantum particle which experiences non-commuting noise in both
the position and momentum direction. For illustrative purposes we treat the simpler
model of this \textit{quantum} stochastic particle in section~\ref{sec:QM noise}.
We calculate the phase space area of a free particle moving under the influence
of two types of noise: \textit{classical white noise} and \textit{instant quantum noise}.
The latter is closely related to the Hamiltonian approach
to stochastic inflation which we discuss in section~\ref{sec:freefieldentropy}
for a free scalar field. By calculating the phase space area
we confirm that a pure state remains pure after the Starobinsky coarse
graining. In section~\ref{sec:selfintpotential} we include a self-interacting
potential for the coarse grained scalar field.
We compare the stochastic Hamilton
equations for a self-interacting field theory to the Starobinsky
equation. A numerical method is proposed for solving the stochastic Hamilton equations
which may present a first quantitative calculation of cosmological
decoherence. Finally in appendices \ref{sec:freefieldquantization} and \ref{sec:QMWigner}
we give basics of the quantisation procedure and the Wigner function approach
to stochastic inflation, respectively.

\section{Classical white noise versus instant quantum noise}
\label{sec:QM noise}

In this section we study how the phase space area of a quantum particle
changes under the influence of environmental noise by considering a simple toy
model of a free quantum mechanical particle where both the position
and momentum of the particle experience some environmental noise. These
two noises induce kicks in the position and momentum direction.
The Hamilton equations for a quantum particle
moving under the influence of the noise terms are
\al{
\dot{{x}}& = \frac{{p}}{m}+F_1(t)\label{Hameqx}\\
\dot{{p}}&=-m\omega^2 {x}+ F_2(t).\label{Hameqp}
}
The time dependent noise terms $F_1(t)$ and $F_2(t)$ are environmental (white) noise,
characterized by the noise correlators
\al{
\langle F_i(t) \rangle &=0\\
\langle F_i(t) F_j(t') \rangle &= f_{ij}(t) \delta(t-t'),
~~~~~~~\{i,j\}=1,2.
\label{noisecorrelators1}
}
The noise $F_2$ is the ordinary noise term for a classical Brownian particle.
Physically it can be viewed as a (big) particle that exchanges small bits of momentum
by moving through an environment of smaller particles. The other noise term $F_1$ has
no classical analogue, as it represents a small infinitesimal displacement of the particle.
However, it allows us to more easily draw an analogy with the stochastic
scalar field in the Starobinsky picture. We will come back to this in the next section.
By substituting Eq. \eqref{Hameqx} into Eq. \eqref{Hameqp} we find an equation for $x$
\eq{
\ddot{x}+\omega^2 x = \dot{F_1}+\frac{1}{m}F_2
\,.
\label{sSHO:eom}
}
We can now solve for $x$ and $p$ by considering the righthand side of \eqref{sSHO:eom} as a source term.
The full solutions of Eqs.~(\ref{Hameqx}--\ref{Hameqp}) are then
\al{
x(t)&=x_0(t)+\int dt' G_{\text{ret}}(t,t')\left[\partial_{t'}F_1(t')+\frac{1}{m}F_2(t')\right]\label{gensolx}\\
p(t)&=p_0(t)+m \int dt' \partial_t G_{\text{ret}}(t,t')\left[\partial_{t'}F_1(t')+\frac{1}{m}F_2(t')\right]-mF_1(t)\label{gensolp},
}
where $x_0(t)$ is the homogeneous solution of Eq.~\eqref{sSHO:eom}, $p_0(t)=m\dot{x}_0(t)$ and the retarded Green's function is
\al{
G_{\text{ret}}(t,t')=\theta(t-t') \frac{\sin [\omega(t-t')]}{\omega}.\label{retgreenx}
}
So far we have not yet discussed canonical quantisation. The position $x$ and momentum $p$
are quantum operators which satisfy the canonical commutator $[x,p]=i$ (setting $\hbar=1$).
There are two ways to implement the quantum character of $x$ and $p$
into the solutions of the Hamilton equations (\ref{gensolx}--\ref{gensolp}):
in \textit{Case 1} the homogeneous solutions $x_0$ and $p_0$ contain the quantum nature of $x$ and $p$.
The noise terms $F_1$ and $F_2$ are consequently \textit{classical white noise} terms.
In \textit{Case 2} the quantumness of $x$ and $p$ originates from the noise terms,
in the sense that $F_1$ and $F_2$ do not commute, \textit{i.e} $[F_1,F_2]\neq 0$.
In order to satisfy the canonical commutator the noise terms have to be of the special type of
\textit{instant quantum noise}.

In both \textit{Case 1} and \textit{Case 2} we are interested in
the phase space area $\Delta$ of the quantum particle,
\eq{
\Delta^2=4\left[\langle x(t)^2 \rangle\langle p(t)^2 \rangle-\langle \frac12\{x(t),p(t)\} \rangle^2\right].
\label{phasespaceareaQM}
}
The phase space area can be used to calculate the Gaussian von Neumann entropy of the system
(see for example Refs. \cite{Koksma:2010zi,Koksma:2010dt}),
\eq{
S_{\text{g}}=\frac{\Delta+1}{2}\ln\left(\frac{\Delta+1}{2}\right)
-\frac{\Delta-1}{2}\ln\left(\frac{\Delta-1}{2}\right).
\label{gaussianvonNeumannentropy}
}
For a pure state the phase space area is minimal, \textit{i.e.} $\Delta=1$,
and consequently the entropy is zero.
In the case of a free quantum mechanical particle without environmental noise
the phase space area is minimal as there are no (environmental or internal)
interactions that could lead to a growth of $\Delta$. On the other hand,
the phase space area in general increases once we include environmental noise.
Our goal is to find out how the phase space area grows in both \textit{Case 1} with
classical white noise and \textit{Case 2} with instant quantum noise.
We will now consider these two cases in more detail.

\subsection{Case 1: Classical white noise}

In the first case, the homogeneous solutions in Eqs.~(\ref{gensolx}--\ref{gensolp}) are
the usual quantum harmonic oscillator solutions expressed in terms of creation and annihilation operators.,
\al{
x_0(t)&= \sqrt{\frac{1}{2 m \omega}}\left({\rm e}^{-i\omega t} {a}
+{\rm e}^{ i \omega t} {a}^{\dagger}\right)
\label{homsolx}
\\
p_0(t)&= i \sqrt{\frac{m \omega}{2}}\left(-{\rm e}^{-i\omega t} {a}
+ {\rm e}^{i \omega t} {a}^{\dagger}\right)
\label{homsolp},
}
with $[a,a^{\dagger}]=1$ and $[a,a]=[a^{\dagger},a^{\dagger}]=0$. These solutions
satisfy the commutation relation $\left[x_0,p_0\right]=i$. Provided that
the noise terms $F_1$ and $F_2$ are classical, \textit{i.e.} commuting, the full solutions
$x$ and $p$ in (\ref{gensolx}--\ref{gensolp}) will satisfy the canonical commutator $\left[x,p\right]=i$.
Now we can calculate the field correlators. The environmental noise and the free particle solutions
are uncorrelated, \textit{i.e.} $\langle x_0 F_i\rangle=\langle p_0 F_i\rangle=0$.
Using Eqs. (\ref{gensolx}--\ref{retgreenx}) and Eq. \eqref{noisecorrelators1} we find
\al{
\langle x(t)^2\rangle&=\frac{1}{2m\omega}+\int^tdt'
\left[c_{\theta}^2 f_{11}(t')+s_{\theta}^2 \frac{f_{22}(t')}{m^2\omega^2}
+s_{\theta}c_{\theta}\frac{\left(f_{12}(t')+f_{21}(t')\right)}{m\omega}\right]\label{x2corr}\\
\langle p(t)^2\rangle&=\frac{m\omega}{2}+m^2\omega^2\int^tdt'
\left[s_{\theta}^2 f_{11}(t')+c_{\theta}^2 \frac{f_{22}(t')}{m^2\omega^2}
-s_{\theta}c_{\theta}\frac{\left(f_{12}(t')+f_{21}(t')\right)}{m\omega}\right]\label{p2corr}\\
\Big\langle \frac12\left\{x(t),p(t)\right\}\Big\rangle
&=m\omega\int^tdt' \left[-c_{\theta}s_{\theta} f_{11}(t')+c_{\theta}s_{\theta} \frac{f_{22}(t')}{m^2\omega^2}+
\frac12(c_{\theta}^2-s_{\theta}^2)\frac{\left(f_{12}(t')+f_{21}(t')\right)}{m\omega}\right]\label{xpcorr},
}
where $s_{\theta}\equiv\sin \theta$, $c_{\theta}\equiv\cos\theta$ and $\theta\equiv \omega(t-t')$.
We take the noise to be classical white noise. This means that the noise correlators in Eq. \eqref{noisecorrelators1}
are constant and that $[F_1,F_2]=0$, \textit{i.e.}
\eq{
f_{ij}(t)=f_{ij}~,~~~~~~~~~f_{12}-f_{21}=0,~~~~~~~~~~~~(\textit{classical white noise}).
\label{quantumbrowniannoise}
}
If we substitute these noise correlators \eqref{quantumbrowniannoise} in the correlators Eqs.~(\ref{x2corr}--\ref{xpcorr})
we find the familiar result for ordinary Brownian motion that at late times
$\langle x(t)^2\rangle \propto t$. Furthermore the variance of the momentum $\langle p(t)^2\rangle \propto t$.
The phase space area \eqref{phasespaceareaQM} of the state becomes
\al{
\nonumber \Delta^2
&=1+2m\omega \left(f_{11}+\frac{f_{22}}{m^2\omega^2}\right)t
\\
&+m^2\omega^2
\left\{\left(f_{11}+\frac{f_{22}}{m^2\omega^2}\right)^2t^2
-\frac{\sin^2\omega t}{\omega^2}
\left[\left(f_{11}-\frac{f_{22}}{m^2\omega^2}\right)^2
      +\frac{\left(f_{12}+f_{21}\right)^2}{m^2\omega^2}
\right]
\right\}.
\label{phasespaceBrowniannoise}
}
At late times this implies
\eq{
\Delta \simeq 1+ m\omega \left(f_{11}+\frac{f_{22}}{m^2\omega^2}\right)t.
}
We clearly see that the phase space area of the initially pure state
eventually grows linearly in time. As a consequence the late time entropy
increases as $S\sim \ln(t)$. These results are all well known for classical
Brownian motion. The difference for the quantum Brownian particle is that
the momentum as well as the position of the particle experiences
a random environmental noise. As a result, the increase in phase space
area depends equally on both noise correlators.\\
To draw a connection with classical Brownian motion, we present here the original Langevin
equation for classical Brownian motion,
\eq{
m\ddot{x}+\zeta \dot{x}=F(t),\label{classicalBrownianmotionLangevin}
}
where $m$ is the mass of the particle and $\zeta$ is a friction coefficient.
The noise terms satisfy
\eq{
\langle F(t)\rangle=0, ~~~~~~~~~~~~~\langle F(t) F(t') \rangle = 2 \zeta k_B T \delta(t-t')\label{Browniancorrelators},
}
with $k_B$ the Boltzmann constant and $T$ the temperature of the particle's environment.
The prefactor on the right-hand side of Eq. \eqref{Browniancorrelators} is fixed by
the fluctuation-dissipation relation. The classical Langevin equation
\eqref{classicalBrownianmotionLangevin} gives the familiar result that
the displacement at late times is proportional to $\sqrt{t}$,
specifically $\langle (x-x_0)^2\rangle\simeq 2 D t$, where $D=k_B T/\zeta$ is
the Einstein diffusion constant. On the other hand the variance of the momentum is
$\langle p^2 \rangle =m^2\langle \dot{x}^2\rangle\simeq m k_BT$. This is to be contrasted
with the variance of the momentum for our model (\ref{Hameqx}--\ref{Hameqp}), where
we find from Eq. \eqref{p2corr} that $\langle p^2 \rangle \propto t$. The difference
originates from the friction term in the classical Langevin equation \eqref{classicalBrownianmotionLangevin}
which limits the growth of $\langle p^2 \rangle$.
Technically the friction term is always present in a thermal bath
because a fluctuation-dissipation relation must be satisfied for a Brownian particle.

\subsection{Case 2: Instant quantum noise \label{sec:instantnoise}}

In the second case the quantumness of $x$ and $p$ in the general solutions (\ref{gensolx}--\ref{gensolp})
originates from the noise terms $F_1$ and $F_2$. The homogeneous solutions are the usual classical solutions
for a free particle
\al{
x_0(t)&=x_{\rm{in}}\cos[\omega(t-t_{\rm in})]+\frac{p_{\rm{in}}}{m\omega}\sin[\omega(t-t_{\rm in})]\label{classhomsolx}\\
p_0(t)&=-m\omega x_{\rm{in}}\sin[\omega(t-t_{\rm in})]+\frac{p_{\rm{in}}}{\omega}\cos[\omega(t-t_{\rm in})]\label{classhomsolp},
}
where $x_{\rm in}$ and $v_{\rm in}$ are the position and velocity of the particle at the initial time $t_{\rm in}$.
Now, using the complete solutions (\ref{gensolx}--\ref{gensolp}) we find that the canonical commutator is
\eq{
[x(t),p(t)]=\int^t dt'\left[ f_{12}(t')-f_{21}(t')\right].
}
This suggest that our noise correlators \eqref{noisecorrelators1}
act only at one instant in time, and that the noise terms commute in a specific way, \textit{i.e.}
\eq{
f_{ij}(t)=f_{ij}\delta(t-t_N),~~~~~~~~~f_{12}-f_{21}=i,~~~~~~~~~~~~~~(\rm{\textit{instant quantum noise}}),
\label{instantnoise}
}
where $t_N$ is the time at which the noise acts. Physically, one can view this
is a classical particle moving freely, then being kicked by
an infinitesimal sheet of smaller particles at time $t_N$.
For classical Brownian motion with instant noise,
the observer will see that the particle receives a small kick of momentum.
On the other hand in the quantum mechanical case with noise in position and momentum (\ref{Hameqx}--\ref{Hameqp}),
the particle experiences a kick of both its momentum and its position,
\textit{i.e.} a \textit{quantum} kick.
As a consequence of this kick, the phase space area of the particle becomes
\footnote{When $\langle x \rangle \neq 0$ and/or $\langle p \rangle \neq 0$,
the correct definition of the phase space area is the following generalisation of \eqref{phasespaceareaQM}:
$$
\Delta^2=4\left[\langle \Delta x^2\rangle\langle \Delta p^2\rangle-\langle \frac12\{\Delta x, \Delta p \} \rangle^2\right],
$$
where $\Delta x =x-\langle x\rangle$ and $\Delta p=p-\langle p\rangle$.
Thus we must subtract any expectation value of $x$ and $p$, which in this case are
the classical homogeneous solutions $x_0$ and $p_0$ in (\ref{classhomsolx}--\ref{classhomsolp}).}
\eq{
\Delta^2=
  4\theta(t-t_N)\left[f_{11}f_{22}-\left(\frac{f_{12}
   + f_{21}}{2}\right)^2\right]
\,.
\label{phasespaceInstantnoise}
}
\begin{figure}
  \center
  \includegraphics[width=.6\textwidth]{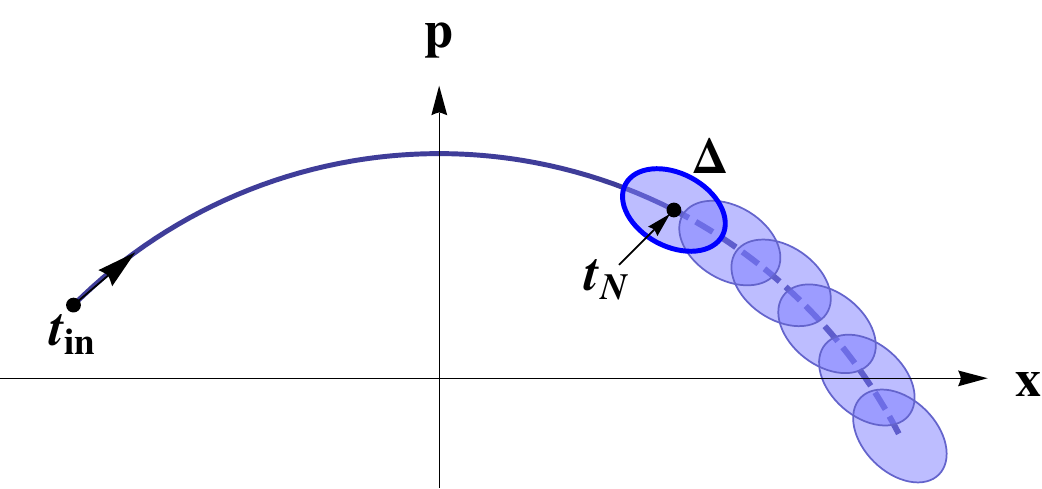}\\
  \caption{\small{Trajectory of a particle influenced by instant quantum noise.
  From time $t_{\rm in}$ the particle follows the classical trajectory of a free particle in~(\ref{classhomsolx}--\ref{classhomsolp}).
  At time $t_N$ the quantum noise kicks in and the particle becomes aware of its quantumness. After $t_N$
  the phase space area $\Delta$ is nonzero as described in \eqref{phasespaceInstantnoise}.
  The classical trajectory is still shown as a dashed line. }}\label{fig:instantnoise}
\end{figure}
Physically the result can be explained in the following way (see Fig.~\ref{fig:instantnoise}):
at a time $t<t_N$ the particle follows a classical trajectory in phase space described by
the classical homogeneous solutions (\ref{classhomsolx}--\ref{classhomsolp}).
The phase space area, expressed in units of $\hbar$, is a typical quantum mechanical
object. For a classical state the phase space area is zero, which can easily be seen
by substituting the classical solutions (\ref{classhomsolx}--\ref{classhomsolp})
into Eq. \eqref{phasespaceareaQM}.

Now, at the time $t_N$ the noise instantly kicks in. Through the noise the particle
suddenly becomes aware of its quantumness, and its phase space area becomes nonzero.
In this sense the instant quantum noise provides a way of writing quantum mechanics
as a stochastic theory. For a consistent quantum mechanical nature of the particle, we find
the condition $4\left[f_{11}f_{22}-\left(\frac{f_{12}
   + f_{21}}{2}\right)^2\right]\geq 1$, such that entropy is well defined.

After the time $t_N$ the phase space area will generally be nonzero.
Consequently, the entropy increases by some amount due to
the quantum kick of the particle. The {\it Case 1} of quantum white noise
means that the particle continuously receives quantum kicks.
As we have seen, the phase space area grows linearly in time in that case.

A special situation occurs when $4\left[f_{11}f_{22}-\left(\frac{f_{12}
   + f_{21}}{2}\right)^2\right]= 1$. In that case the phase space area is minimal
and the initially classical particle is now in a pure quantum mechanical state.
As we will see in the next section, in stochastic inflation~\cite{Starobinsky:1986bb}
the modes for a noninteracting field exhibit
a similar behavior. In the phase space approach to stochastic inflation
instant quantum noise is present in the Hamilton equations
for the super-Hubble, or infrared (IR) modes of the inflaton field.
The noise appears because of a special type of coarse graining,
which is based on integrating out the sub-Hubble,
or ultraviolet (UV) modes of the field. The properties of the noise are however
extremely special, in the sense that the ultraviolet modes that cross the horizon
are in a pure state, thus there is no entropy generation.
We will examine this now in greater detail.

\section{Correlators and entropy for free coarse-grained scalar field}
\label{sec:freefieldentropy}

We start with a field theory of a free massless scalar field in an expanding
universe. Details of the theory and the quantized solutions of the fields
and momenta can be found in Appendix~\ref{sec:freefieldquantization}.
In the phase space approach to stochastic inflation~\cite{Habib:1992ci}
the scalar field~\eqref{fieldgeneralsolution} and
momentum~\eqref{momentumgenralsolution} are separated into super- and
sub-Hubble modes, or equivalently IR and UV modes,
\al{
\Phi&=\phi+\varphi\label{stochasticsplittingphi}\\
\Pi&=\pi_{\phi}+\pi_{\varphi}\label{stochasticsplittingpi}.
}
The super-Hubble modes $\phi$ and $\pi_{\phi}$ of the scalar field and
momentum are those modes with $k\ll Ha$. We separate these modes from the
rest by introducing a time dependent cut-off $\theta(\epsilon aH-k)$, where
$\epsilon \lesssim 1$, and
\al{
\phi(t,\bvec{x})&=\int \frac{d^{3}\bvec{k}}{(2\pi)^{3}}\theta(\epsilon aH-k)
\left\{u(t,k){\rm e}^{i\bvec{k}\cdot\bvec{x}}{\alpha}(\bvec{k})
+ u^{\star}(t,k){\rm e}^{-i\bvec{k}\cdot\bvec{x}}{\alpha}^{\dagger}(\bvec{k})
\right\}\label{coarsegrainedphi}\\
\pi_{\phi}(t,\bvec{x})&=a^{3}\int \frac{d^{3}\bvec{k}}{(2\pi)^{3}}
\theta(\epsilon aH-k)
\left\{\dot{u}(t,k){\rm e}^{i\bvec{k}\cdot\bvec{x}}{\alpha}(\bvec{k})
+ \dot{u}^{\star}(t,k){\rm e}^{-i\bvec{k}\cdot
        \bvec{x}}{\alpha}^{\dagger}(\bvec{k})
\right\}
\label{coarsegrainedpi}.
}
Similarly, the short wavelength expressions $\varphi$ and $\pi_{\varphi}$
are the other modes with $k>\epsilon Ha$. The time dependent momentum
cut-off in the short wavelength terms leads to a stochastic noise term in the
long wavelength equations. We can easily see this by substituting
Eqs.~(\ref{stochasticsplittingphi}--\ref{stochasticsplittingpi}) into
the Hamilton equations~(\ref{Hamiltoneqdotphi}--\ref{Hamiltoneqdotpi})
and using the field equation for the mode functions~\eqref{modefunctionseq}.
We find
\al{
\dot{\phi}&=\frac{1}{a^3}\pi_{\phi}+F_1(t,\bvec{x})
\label{stochHamiltoneqphi}
\\
\dot{\pi}_{\phi} &=a^{3}\frac{\nabla^2}{a^2}\phi+F_2(t,\bvec{x})
\label{stochHamiltoneqpi}
\,,
}
where the stochastic noise terms are
\al{
F_1(t,\bvec{x})&=\int \frac{d^{3}\bvec{k}}{(2\pi)^{3}}\delta(k-\epsilon aH)
(\epsilon aH)^{\cdot}
\left\{u(t,k){\rm e}^{i\bvec{k}\cdot\bvec{x}}{\alpha}(\bvec{k})
+ u^{\star}(t,k){\rm e}^{-i\bvec{k}\cdot\bvec{x}}{\alpha}^{\dagger}(\bvec{k})
\right\}
\label{stochnoiseF1}
\\
F_2(t,\bvec{x})&=a^{3}\int \frac{d^{3}\bvec{k}}{(2\pi)^{3}}
\delta(k-\epsilon aH)(\epsilon aH)^{\cdot}
\left\{\dot{u}(t,k){\rm e}^{i\bvec{k}\cdot\bvec{x}}{\alpha}(\bvec{k})
+ \dot{u}^{\star}(t,k){\rm e}^{-i\bvec{k}
       \cdot\bvec{x}}{\alpha}^{\dagger}(\bvec{k})
\right\}
\,.
\label{stochnoiseF2}
}
where $u(t,k)$ are the scalar field mode functions defined
in~(\ref{fieldgeneralsolution}--\ref{momentumgenralsolution}),
and $\alpha(\bvec{k})$ and $\alpha^\dagger(\bvec{k})$ are the annihilation
and creation operators~(\ref{creanncommutator}). Note that the stochastic Hamilton equations
in the field theoretical case~(\ref{stochHamiltoneqphi}--\ref{stochHamiltoneqpi}) are (almost) the same
as those in the quantum mechanical case~(\ref{Hameqx}--\ref{Hameqp}).
At first it was believed that the coarse-graining of the free field
could lead to decoherence due to the time dependent split
in long and short wavelength modes~\cite{Hosoya:1988yz,Nambu:1991vs}.
Habib~\cite{Habib:1992ci} however argued that the stochastic inflation
type of coarse-graining for a free field
does not lead to decoherence. Here we will put these arguments on
a firmer footing by performing an explicit calculation of (potential)
decoherence for coarse-grained free scalar fields.
We will do this in a general accelerating FLRW universe without assuming
a specific choice of the scale factor.

Our first goal is to show that
coarse-graining through a time dependent cutoff does not lead to a growth of
the phase space area~\eqref{phasespacearea} and Gaussian entropy~\eqref{EntropyvonNeumann}
for the super-Hubble modes of
the scalar field. In order to do so we will be interested in
the momentum space correlators for $\phi$ and $\pi_{\phi}$ as they
determine the phase space area for every mode $k$. In principle
we can proceed along the same lines as the quantum mechanical particle with
quantum white noise in section~\ref{sec:QM noise}. The (free) scalar field
is namely nothing more than an infinite sum of simple harmonic oscillators.
Analogously to~(\ref{sSHO:eom}), by combining~(\ref{stochHamiltoneqphi})
and~(\ref{stochHamiltoneqpi}) and transforming into a Fourier space,
we can write the field equations for the modes
$\phi(t,\bvec{k})$ and $\pi_{\phi}(t,\bvec{k})$:
\al{
\ddot{\phi}(t,\bvec{k})+3H\dot{\phi}(t,\bvec{k})
 + \frac{k^2}{a^2}\phi(t,\bvec{k})
=\frac{1}{a^{3}}\left[(a^{3}F_{1}(t,\bvec{k}))^{\centerdot}
+F_{2}(t,\bvec{k})\right]
\label{stochfieldequation}
}
and solve for $\phi(t,\bvec{k})$ using the method of Green's functions like
in Eqs.~\eqref{gensolx} and~\eqref{gensolp}.
However, there is an important difference with respect to the quantum mechanical case.
We obtained the stochastic Hamilton equations~(\ref{stochHamiltoneqphi}--\ref{stochHamiltoneqpi})
by coarse graining the homogeneous solutions~(\ref{fieldgeneralsolution}--\ref{momentumgenralsolution})
and substituting the coarse grained field
and momentum~(\ref{coarsegrainedphi}--\ref{coarsegrainedpi}) back into the original Hamilton
equations~(\ref{Hamiltoneqdotphi}--\ref{Hamiltoneqdotpi}).
It would therefore make no sense to add the homogeneous solution of
Eq.~\eqref{stochfieldequation} to the coarse-grained solution.
Physically speaking the homogeneous solution is absent, as the IR quantum field
only exists once the UV field has crossed the Hubble radius, which is described
through the quantum noise terms (\ref{stochnoiseF1}--\ref{stochnoiseF2}) alone.

Mathematically speaking, we could solve Eqs.~(\ref{stochHamiltoneqphi}--\ref{stochHamiltoneqpi})
without knowing the origin of the equations,
but with knowledge of the noise terms (\ref{stochnoiseF1}--\ref{stochnoiseF2}).
Because the quantumness of $\phi$ and $\pi_{\phi}$ is
contained in the noise terms $F_1$ and $F_2$, we could
still add a classical homogeneous solution.
However, as we have also seen in section \ref{sec:instantnoise}, the phase space area
is defined in terms of quantum fluctuations on top of the classical solution. Therefore
any classical solution does not appear in the phase space area and can for our purposes be neglected.
Thus, the full (quantum) solutions for $\phi(t,\bvec{k})$ and
$\pi_{\phi}(t,\bvec{k})$ are just
\al{
\phi(t,\bvec{k})&=\int dt'G_{\text{ret}}(t,t',k)\left[(a^{\prime 3}
        F_{1}(t',\bvec{k}))^{\centerdot}+F_{2}(t',\bvec{k})\right]
\label{stochGreensolphi}\\
\pi_{\phi}(t,\bvec{k})&=a(t)^3\int dt'
    \left[\partial_t G_{\text{ret}}(t,t',k)\right]
  \left[(a^{\prime 3}F_{1}(t',\bvec{k}))^{\centerdot}\
            + F_{2}(t',\bvec{k})\right]-a(t)^3F_{1}(t,\bvec{k})
\,,
\label{stochGreensolpi}
}
where $a'\equiv a(t')$. The retarded Green's function is given in
Eq.~\eqref{retardedGreensfk} of Appendix~\ref{sec:freefieldquantization}.
Of course, if one substitutes the actual noise terms~(\ref{stochnoiseF1}--\ref{stochnoiseF2}) into the above solutions,
we will recover again the coarse grained solutions in
Eqs.~(\ref{coarsegrainedphi}--\ref{coarsegrainedpi}).
At this moment however we will keep using the solutions for
$\phi(t,\bvec{k})$ and $\pi_{\phi}(t,\bvec{k})$ in terms of the noise terms.
Here we wish to calculate the phase space area $\Delta_k$ in terms of
the noise correlators. This allows us to draw an analogue with
the quantum mechanical case in the previous section.
First let us calculate the correlators for the noise terms in momentum space,
which we can write in the following way,
\eq{
\langle F_{i}(t,\bvec{k})F^{\ast}_{j}(t',\bvec{k}) \rangle
   = f_{ij}\delta(t-t_H)\delta(t-t')
\,,
\label{stochnoisecorrelators}
}
with
\al{
f_{11}&=|u(t_H,k)|^2\\
f_{22}&=a^6(t_H)|\dot{u}(t_H,k)|^2\\
\frac12 \left(f_{12}+f_{21}\right)&=
\frac12 a^3(t_H)\Bigl(u(t_H,k)\dot{u}^{\star}(t_H,k)+\dot{u}(t_H,k)u^{\star}(t_H,k)\Bigr).\label{stochnoisecorrelatorsfij}
}
In these expressions $t_H$ is the time of the mode split, \textit{i.e.} the instant of time when
$k=\epsilon Ha$. From these expressions one can also see that the noise terms
$F_1$ and $F_2$ do not commute, since
\eq{
[F_{1}(t,\bvec{k}),F^{\ast}_{2}(t',\bvec{k})]
 = (f_{12}-f_{21}) \delta(t-t_H)\delta(t-t')=i\delta(t-t_H)\delta(t-t')
\,,
\label{noisecommutator}
}
such that, strictly speaking, $F_{1}$ and  $F_{2}$ are operator valued noises.
We have used here the Wronskian condition~\eqref{Wronskiancondition}.
The crucial observation of the noise correlators in
Eq.~\eqref{stochnoisecorrelators} is that they are precisely of
the~\textit{instant quantum noise} type that we discussed in
section~\ref{sec:instantnoise}, see Eq. \eqref{instantnoise}. With this knowledge we can now
calculate the phase space area~\eqref{phasespacearea}
for the solutions~(\ref{stochGreensolphi}--\ref{stochGreensolpi}).
After various partial integrations and
use of the Wronskian condition, we find
\al{\Delta_k^2
 &= 4\theta(t-t_H)\left[f_{11}f_{22}
   - \left(\frac{f_{12}+f_{21}}{2}\right)^2\right]
\,.
\label{phasespaceareaphi}
}
Notice that $\Delta_k$ does not depend on $f_{12}-f_{21}$ and hence
the operator character of the noises is not relevant for the calculation
of the phase space area and thus neither for the calculation of the Gaussian entropy of the state.
Not surprisingly, the phase space area of the IR modes~(\ref{phasespaceareaphi})
is precisely equal to the phase space area of the free particle influenced
by instant quantum noise~(\ref{phasespaceInstantnoise}). Thus also here,
once the UV modes cross the Hubble radius (\textit{i.e.} at time $t_H$),
the quantumness of the IR modes kicks in through a nonzero phase space area.

The question remains: how big is the phase space area for the super Hubble modes
of an interacting scalar field under the influence of noise coming from the
sub Hubble modes? Let us first consider the free scalar field theory from this section.
In this case we know exactly what are the noise correlators
from Eq.~\eqref{stochnoisecorrelators}. We can therefore simply calculate
the phase space area~\eqref{phasespaceareaphi} of the coarse grained scalar
field by inserting the noise
correlators~(\ref{stochnoisecorrelators}--\ref{stochnoisecorrelatorsfij}).
By using the Wronskian
condition~\eqref{Wronskiancondition} twice we obtain the result
\eq{
\Delta_k^2=\theta(t-t_H)=\theta(\epsilon aH-k)
\,.
\label{resultphasespacefreefield}
}
Thus for the super-Hubble modes with $k<\epsilon Ha$
the phase space area is always exactly $1$, \textit{i.e.} the system
is in a pure state. The Gaussian von Neumann entropy~\eqref{EntropyvonNeumann}
is therefore
\eq{
S_{\text{g}}=\int^{k<\epsilon a H}\!\! \frac{d^{3}\bvec{k}}{(2\pi)^{3}}
\left[\frac{\Delta_k+1}{2}\ln\left(\frac{\Delta_k+1}{2}\right)
  -\frac{\Delta_k-1}{2}\ln\left(\frac{\Delta_k-1}{2}\right)\right]=0
\,.
\label{EntropyvonNeumannresultfreefield}
}
This reflects the general fact that introducing a time dependent mode
separation of a free scalar field by means of the Heaviside step function
does not generate any entropy.
Even though the Starobinsky coarse graining introduces noise terms
for both the field and its conjugate momentum, the properties of
the noise are such that, once the modes cross the Hubble radius, the phase
space area is minimal. The reason is that the noise appears here
not because of some coupling to an external environment, but because
sub-Hubble modes are continuously crossing the Hubble radius.

The same results for the phase space area~\eqref{resultphasespacefreefield}
and entropy~\eqref{EntropyvonNeumannresultfreefield} can be, of course,
obtained by using the coarse-grained solutions~(\ref{coarsegrainedphi}--\ref{coarsegrainedpi}).
We emphasize that these solutions are the exact solutions of the Hamilton equations
with noise (\ref{Hamiltoneqdotphi}--\ref{Hamiltoneqdotpi}).
If we use these solutions from the start, we trivially find
\al{
\langle |\phi(t,\bvec{k})|^2\rangle
     & = \theta(\epsilon aH-k) |u(t,k)|^2
\label{genformphi2}
\\
\langle |\pi_{\phi}(t,\bvec{k})|^2\rangle
  & = \theta(\epsilon aH-k) a^6(t)|\dot{u}(t,k)|^2
\label{genformpi2}
\\
\Big\langle \frac12 \left\{\phi(t,\bvec{k}),\pi^{\ast}_{\phi}(t,\bvec{k})
\right\}\Big\rangle
& = \theta(\epsilon aH-k) a^3(t)
 \Bigl(u(t,k)\dot{u}^{\star}(t,k)+\dot{u}(t,k)u^{\star}(t,k)\Bigr)
\label{genformphipi}
\,.
}
Substituting these propagators into the phase space
area~\eqref{phasespacearea} gives as result again
Eq.~\eqref{resultphasespacefreefield}. The results here are thus general:
once we have a free field and produce noise terms by introducing a
time dependent cut-off for the modes, the entropy of the coarse-grained
field will not grow. Here we considered a massless field, but even if
we add a (time dependent) mass term, in principle we can still find
normalized mode functions $u(t,k)$ and we can express the coarse-grained fields
as in Eqs.~(\ref{coarsegrainedphi}--\ref{coarsegrainedpi}). In general
adding or changing a mass term in the Hamilton equations only changes
the squeezing of a state, not its phase space area.
Besides adding a mass term, we could also add a quartic potential $\lambda \phi^4/4!$
for the coarse-grained fields. In a mean field approach we can write
the corresponding Hamilton equations for the super-Hubble field and momentum as
\al{
\dot{\phi}&=\frac{1}{a^3}\pi_{\phi}+F_1(t,\bvec{x}) \label{stochHamiltoneqphiSC}\\
\dot{\pi}_{\phi}
 &=a^{3}\left(\frac{\nabla^2}{a^2}-m^2
             -\frac{\lambda}{2}\langle \phi^2(t)\rangle \right)\phi
          + F_2(t,\bvec{x})
\label{stochHamiltoneqpiSC}
\,.
}
The term $m^2-\lambda\langle \phi^2(t)\rangle/2$ is
a time dependent mass term,
and hence in a mean field approximation the field remains a free field
and the entropy does not grow.
We emphasize that, in the original Starobinsky's approach,
both interactions between the sub-Hubble fields as well as
interactions between the sub-Hubble and super-Hubble fields
are neglected. One can show~\cite{Tsamis:2005hd} that,
for a scalar field with an arbitrary potential
in the stochastic formulation of Starobinsky, one correctly  captures
all leading order logarithms $[\ln(a)]^n$ of coincident correlators
to all orders in perturbation theory.
The stochastic approach has been extended to
Yukawa theory~\cite{Miao:2006pn} and to
scalar electrodynamics~\cite{Prokopec:2006ue,Prokopec:2007ak,Prokopec:2008gw},
but no stochastic formulation
of quantum gravity is as yet known.


An important question is then what can source decoherence
within stochastic inflation.
The simple answer is: one source of decoherence
are interactions between the infrared fields treated beyond the
mean field approximation. This source is present in
our Hamiltonian formulation of stochastic inflation
and it is the basis for the next section.
There are, of course, other sources of decoherence in the full quantum theory,
the most notable one are interactions between the sub-Hubble and
super-Hubble fields. Including these would require a significant modification
of the framework of stochastic inflation, and will not be considered in this work.
Some aspects of these interactions are studied in Ref.~\cite{Matacz:1996gk}.
However, Matacz (incorrectly) neglects the leading noise terms $F_1$ and
$F_2$~(\ref{stochnoiseF1}--\ref{stochnoiseF2})
that are already present in the free
theory~(\ref{stochHamiltoneqphi}--\ref{stochHamiltoneqpi}).
Finally, we remark that, based on the results of the study conducted in
Ref.~\cite{Koksma:2009wa}, we do not expect interactions between
the sub-Hubble modes to generate any significant entropy.

\section{Correlators and entropy for a self-interacting scalar field}
\label{sec:selfintpotential}

 Our main interest is to study decoherence of scalar cosmological perturbations
during inflation. Here we model this problem by using a toy model (which
captures some, but not all important features of cosmological perturbations):
a self-interacting scalar field with a quartic potential
$V(\phi) = \lambda \phi^4/4!$ in an accelerating universe.
Since we are interested in the phase-space area, a full Hamiltonian analysis
is necessary. The resulting classical stochastic equations are ({\it cf.}
Eq.~(\ref{stochHamiltoneqpiSC})),
\al{
\dot{\phi}&=\frac{\pi_{\phi}}{a^3}+F_1(t,\bvec{x})
\label{stochHamiltoneqphiINT}
\\
\dot{\pi}_{\phi} &=a\nabla^2\phi+F_2(t,\bvec{x})
- a^3\frac{\lambda}{3!}\phi^3
\,.
\label{stochHamiltoneqpiINT}
}
We emphasize that here $F_1$ and $F_2$ are
are {\it not} the quantum noises given in
Eqs.~(\ref{stochnoiseF1}--\ref{stochnoiseF2}), but
they are classical noise terms, which can be obtained
from~(\ref{stochnoiseF1}--\ref{stochnoiseF2})
by replacing the creation and annihilation operators
by classical random variables as follows,
\begin{equation}
\nonumber {\alpha}(\bvec{k})~\rightarrow~\alpha_c(\bvec{k})
\,,\qquad
{\alpha}^{\dagger}(\bvec{k})~\rightarrow~\alpha_c^{\star}(\bvec{k})
\,,
\label{alphaclassical}
\end{equation}
where $\alpha_c^{\star}(\bvec{k})$ is simply the complex conjugate of
$\alpha_c(\bvec{k})$. If the initial state is taken to be pure and
free, then the classical (commuting) random variables
 $\alpha_c(\bvec{k})
   = \Re[\alpha_c(\bvec{k})]+i\Im[\alpha_c(\bvec{k})]$
are drawn from a Gaussian phase space distribution,
which can correspond to a pure state (with the surface area $\Delta = 1$),
which can be squeezed and/or displaced, or to a mixed state
with $\Delta_k>1$ (at least for some $\bvec{k}$).
This prescription resembles in spirit the methods used for
studying preheating after inflation~\cite{Prokopec:1996rr}.
Several comments are now in order.

 Replacing the noises $F_{1,2}$ in~(\ref{stochHamiltoneqphiINT}--\ref{stochHamiltoneqpiINT})
by their classical counterparts
makes sense as long as correlators that involve anticommutators
are large when compared with those that involve commutators of
fields (and their momenta), and this is justified for the super-Hubble
modes (see also Ref. \cite{Tolley:2008na} for a discussion about classicality). The spirit of the classical approximation becomes even clearer when
one uses perturbative two-particle irreducible (2PI) methods to
study out-of equilibrium field dynamics.
From {\it e.g.} Refs.~\cite{Koksma:2009wa} and~\cite{Koksma:2011dy}
it follows that evolution of the statistical Green function
(which involves a field anticommutator) depends also on (an integral over)
the causal Green function (involving a field commutator).
Now, if the causal Green function
is much smaller than the statistical Green function, one can
neglect the former in the equation for the statistical Green function,
resulting in a classical approximation. This classical approximation
is the 2PI analogue of the 1PI classical scheme advocated here.
Ultimately, the validity of our scheme can be tested
by comparing with the full (quantum) 2PI methods, which would require
suitable adaptations of methods in Refs.~\cite{Giraud:2009tn},~\cite{Koksma:2009wa}
and~\cite{Koksma:2011dy}.

 For a quantitative assessment of decoherence in cosmological settings
one needs the Gaussian
von Neumann entropy~(\ref{EntropyvonNeumann}--\ref{phasespacearea}),
which is a function of the phase space area $\Delta$,
and which in turn depends on correlators
that involve only anticommutators. This then implies that solving
the classical equations~(\ref{stochHamiltoneqphiINT}--\ref{stochHamiltoneqpiINT}) should suffice.

 The main purpose of this work was to propose a {\it simple} method
suitable for studying decoherence in cosmological settings.
We postpone an actual numerical investigation to a future paper,
where more emphasis will be given to the important question:
how are cosmological observables influenced by the amount of decoherence
of cosmological perturbations.

 Of course, the stochastic theory~(\ref{stochHamiltoneqphiINT}--\ref{stochHamiltoneqpiINT})
must be discretized in order to solve these equations numerically.
The theory will be discretized in both position and momentum space
at the level of the action in order to keep unitarity of the resulting
(discrete) equations of motion~(\ref{stochHamiltoneqphiINT}--\ref{stochHamiltoneqpiINT}).
This is important, since such a discretization method guarantees that
discretization will not induce a new source of decoherence.

 A further assumption in our stochastic approach is that there are no
interactions between the UV and IR parts of the field. The validity
of this approximation can be tested by making a comparison with
a full 2PI evolution.  This was firstly done by Giraud and
Serreau~\cite{Giraud:2009tn} in Minkowsky space, and what remains to be done
is to extend their work to inflationary spaces.

Finally we note that alternatively one could calculate the Gaussian correlators
in the phase space area from the Wigner distribution. This Wigner distribution can be
solved from a Liouville equation which can be obtained from the stochastic Hamiltonian.
For illustrative purposes we derive the Liouville equation for the toy model
of the stochastic quantum particle in appendix \ref{sec:QMWigner}. Although an alternative
to our proposed method, calculating cosmological decoherence through the Wigner distribution method
is much harder due to the non-local nature of the self-interacting field in momentum space.

\section{Conclusion and Discussion}
\label{Conclusion and Discussion}

 In this work we propose a generalised framework of stochastic inflation
that is suitable for studies of decoherence of scalar cosmological
perturbations during inflation. We take the Gaussian von Neumann entropy
as a quantitative measure of decoherence.
This means that the entropy is generated by
neglecting observationally inaccessible higher order field correlators.

 In sections~\ref{sec:QM noise} and~\ref{sec:freefieldentropy}
free theories are discussed. Section~\ref{sec:QM noise} is mostly pedagogical.
There we first consider the quantum stochastic particle with
(white) noise in the direction of position and momentum. The canonical commutator
can be satisfied by having a free quantum mechanical particle moving under the influence
of classical white noise. In that case the phase space area increases linearly in time at late times.
However, the quantumness of position and momentum can also originate from the noise terms themselves.
The noise in that case is instant quantum noise, with non-commuting noise terms in the position
and momentum direction which only act at one instant of time.
In that case the initially classical particle
receives a quantum kick and becomes aware of its quantumness by having a nonzero phase space area.
The latter is closely related to the stochastic noise present in
the phase space approach to stochastic inflation, of which we discuss the free case in
section~\ref{sec:freefieldentropy}.
The origin of the (quantum) noise
is in fact that the sub-Hubble modes continuously cross the Hubble radius.
It has the special property that it is unitary, \textit{i.e.} once
the modes exit the Hubble radius, the phase space area remains minimal
and constant, implying that the modes are in a pure state.
This special property guarantees that there is no decoherence for a
free stochastic inflationary theory. In addition,
there is no decoherence when interactions are treated in a mean field
approximation.

Next, in section~\ref{sec:selfintpotential}, a classical Hamiltonian
stochastic inflation for a self-interacting scalar field has been proposed,
which should be suitable for studies of decoherence of cosmological
perturbations during inflation. In our approach,
the noises are taken to be classical stochastic commuting variables,
such that the equations of motion are suitable for numerical treatment.
 The main advantage of our approach (when compared to more sophisticated
2PI methods) is its simplicity. Of course, it would be very nice to have
an analytic argument concerning the accuracy of our method. In the absence of it,
we point out that our approach to decoherence can be tested by the more sophisticated 2PI methods.

 Finally, we note that understanding cosmological decoherence
is important, since a highly decohered state of scalar cosmological
perturbations will shift the first acoustic peak
and reduce the amplitude of the secondary peaks in the CMB radiation
(see, for example, Ref.~\cite{Durrer:1997rh}).
The ultimate goal of this project is to find out how strong
the cosmological decoherence is in specific inflationary models,
such as a chaotic $\lambda \Phi^4$ theory,
and to find out what are the observable effects on the CMB.

\section*{Acknowledgements}
We would like to thank Jurjen Koksma for stimulating discussions and for
useful comments on the draft.
We acknowledge support from the Dutch Foundation for 'Fundamenteel Onderzoek der Materie'
(FOM) under the program "Theoretical particle physics in the era of the LHC", program number FP 104.

\appendix

\section{\label{sec:freefieldquantization} Free field quantization}
Here we recall the physics of canonical quantisation of a free scalar field in inflation.
We start with the action of a free scalar field on a general background
\begin{equation}
S_{\text{free}}=-\frac12 \int d^4x \sqrt{-g}g^{\mu\nu}\partial_\mu\Phi\partial_\nu\Phi.
\label{freelagrangian}
\end{equation}
We use the FLRW metric $g_{\mu\nu}=\text{diag}(-1,a^2,a^2,a^2)$ with $a=a(t)$ the scale factor.
The canonical momentum for $\Phi$ is
\eq{
\Pi_{\Phi}=\frac{\partial \mathcal{L}}{\partial \dot{\Phi}} = a^3 \dot{\Phi},
\label{canonicalmomentum}
}
where $\mathcal{L}$ is the Lagrangian density, $S=\int d^4x \mathcal{L}$.
The Hamiltonian density becomes
\eq{
\mathcal{H}= \frac12\left\{ \frac{\Pi_{\Phi}^2}{a^3}+a^3\left(\frac{\nabla\Phi}{a}\right)^2\right\}
\label{freehamiltonian}
}
The free Hamiltonian $H$ is related to the Hamiltonian density as $H=\int d^3x \mathcal{H}$.
The Hamilton equations are
\al{
\dot{\Phi}&=\frac{\partial\mathcal{H}}{\partial\Pi_\Phi}=\frac{1}{a^3}\Pi_\Phi \label{Hamiltoneqdotphi} \\
\dot{\Pi}_{\Phi}&=-\frac{\partial\mathcal{H}}{\partial\Phi}=a^{3}\frac{\nabla^2}{a^2}\Phi \label{Hamiltoneqdotpi}.
}
Combining these two equations gives the free field equation
\eq{
\ddot{\Phi}+3 H\dot{\Phi}-\frac{1}{a^2}\nabla^2\Phi=0.
\label{freefieldequation}
}
We now quantize the system by imposing the canonical commutator
\eq{
[\Phi(t,\bvec{x}),\Pi_\Phi(t,\bvec{x}')]=i\delta^{3}(\bvec{x}-\bvec{x}').
\label{canonicalcommutator}
}
A general solution to the Hamilton equations~\eqref{Hamiltoneqdotphi}
and~\eqref{Hamiltoneqdotpi} which satisfies this canonical commutation
relation is
\al{
\Phi(t,\bvec{x})&=\int \frac{d^{3}\bvec{k}}{(2\pi)^{3}}
\left\{u(t,k)e^{i\bvec{k}\cdot\bvec{x}}{\alpha}(\bvec{k})
 + u^{\star}(t,k)e^{-i\bvec{k}\cdot\bvec{x}}{\alpha}^{\dagger}(\bvec{k})
 \right\}
\label{fieldgeneralsolution}
\\
\Pi_{\Phi}(t,\bvec{x})&=a^{3}(t)\int \frac{d^{3}\bvec{k}}{(2\pi)^{3}}
\left\{\dot{u}(t,k)e^{i\bvec{k}\cdot\bvec{x}}{\alpha}(\bvec{k})
+\dot{u}^{\star}(t,k)e^{-i\bvec{k}\cdot\bvec{x}}{\alpha}^{\dagger}(\bvec{k})
\right\}.
\label{momentumgenralsolution}
}
The creation and annihilation operators satisfy the commutation relations
\eq{
[{\alpha}(\bvec{k}),{\alpha}^{\dagger}(\bvec{k}')]
 =(2\pi)^{3}\delta^{3}(\bvec{k}-\bvec{k}')
\,,\qquad
[{\alpha}(\bvec{k}),{\alpha}(\bvec{k}')]  = 0
\,,\qquad
[{\alpha}^{\dagger}(\bvec{k}),{\alpha}^{\dagger}(\bvec{k}')] = 0
\,.
\label{creanncommutator}
}
The mode functions $u(t,k)$ obey the free field equation in momentum space
\eq{
\ddot{u}+3H\dot{u}+\frac{k^2}{a^2}u=0,
\label{modefunctionseq}
}
and satisfy the Wronskian normalization condition
\eq{
W[u(t,k),u^{\star}(t,k)]=u(t,k)\dot{u}^{\star}(t,k)-u^{\star}(t,k)\dot{u}(t,k)=\frac{i}{a^{3}}.
\label{Wronskiancondition}
}
The retarded Green's function for the field is
\al{
\nonumber G_{\text{ret}}(t,\bvec{x};t',\bvec{x}') & = i\theta(t-t') \langle \left[ \Phi(t,\bvec{x}),\Phi(t',\bvec{x}')\right]\rangle\\
& =i\theta(t-t')\int  \frac{d^{3}\bvec{k}}{(2\pi)^{3}}e^{i\bvec{k}\cdot \Delta\bvec{x}}\left\{u(t,k)u^{\star}(t',k)-u^{\star}(t,k)u(t',k)\right\}.
\label{retardedGreensf}
}
which satisfies
\eq{
\left(\partial_t^2+3 H \partial_t-\frac{1}{a^2}\nabla^2\right)G_{\text{ret}}(t,\bvec{x};t',\bvec{x}')=\delta(t-t')\frac{\delta^3(\bvec{x}-\bvec{x}')}{a^3}.
}
Its Fourier transform is
\eq{
G_{\text{ret}}(t,t',k)=i\theta(t-t')\left\{u(t,k)u^{\star}(t',k)-u^{\star}(t,k)u(t',k)\right\}.
\label{retardedGreensfk}
}

\section{\label{sec:QMWigner} Phase space area in stochastic quantum mechanics}
An alternative approach to the calculation of the phase space area and entropy of
stochastic fields is to solve the quantum Liouville equation. Here we focus on a
simple quantum mechanical example, but the approach can in principle be extended
to coarse-grained inflaton fields. We will follow mostly the approach by Habib
\cite{Habib:1992ci}, which is in turn based on Kubo's approach \cite{Kubo:1963}.
We will clarify where necessary.\\
We start with the stochastic quantum mechanical equations of motion,
\al{
\dot{x}& = \frac{p}{m}+F_1\\
\dot{p}&=-m\omega^2 x-V'({x})+F_2,
}
which in principle follow from the Hamiltonian
\eq{
{H}=\frac{1}{2m}{p}^2+\frac12 m\omega^2 {x}^2+V({x})+F_1{p}-F_2{x}.
}
$x$ and $p$ obey  the canonical commutation relation $[x,p]=i\hbar$.
The time dependent noises satisfy
\al{
\langle F_i(t) \rangle &=0\\
\langle F_i(t_1) F_j(t_2) \rangle &= f_{ij}(t_1) \delta(t_1-t_2),~~~~~~~\{i,j\}=1,2.\label{noisecorrelators}
}
The density operator $\hat{\rho}$ satisfies the von Neumann equation (the quantum Liouville equation)
\eq{
i\hbar \partial_t {\rho}=\left[{H},{\rho}\right].
}
In position space the density operator can be represented as
\eq{
{\rho}=\int_{-\infty}^{+\infty}dx \int_{-\infty}^{+\infty}dx' |x\rangle \rho(x,x';t)\langle x'|.
}
The von Neumann equation projected onto position space becomes
\al{
\nonumber i\hbar \partial_t \rho(x,x';t)&=\Biggl\{-\frac{\hbar^2}{2m}
(\partial_x^2-\partial_{x'}^2) +\frac{m\omega^2}{2}(x^2-x'^2)+V(x)-V(x')\\
&-i\hbar F_1(\partial_x+\partial_{x'})-F_2(x-x')\Biggr\}\rho(x,x';t)\label{vonNeumannequationQM}
}
where we have made used that ${p}|x\rangle=-i\hbar \partial_x |x\rangle$
using the canonical commutator  $[{x},{p}]=i\hbar$. Now we perform a Wigner
transform of the density matrix. This gives the Wigner function
\eq{
\mathcal{W}(q,p;t)=\int_{-\infty}^{\infty}d\Delta{x}e^{-i p \Delta x/\hbar}\rho\left(q+\frac{\Delta x}{2},q-\frac{\Delta x}{2};t\right).
\label{Wignertransform}
}
The new coordinates are defined as
\al{
q&=\frac{x+x'}{2}\\
\Delta x&= x-x'.
}
Substituting these coordinates into the von Neumann equation \eqref{vonNeumannequationQM} and using that
\al{
(\partial_x^2-\partial_{x'}^2)\rho(x,x';t)&=2\partial_q\partial_{\Delta x}\rho\left(q+\frac{\Delta x}{2},q-\frac{\Delta x}{2};t\right)\\
\Delta x e^{-i p \Delta x/\hbar}&= i\hbar \left(\partial_pe^{-i p \Delta x/\hbar}\right),
}
we find that the equation for the Wigner function \eqref{Wignertransform} becomes
\eq{
\left\{\partial_t+\frac{p}{m}\partial_q-m\omega^2q\partial_p+\frac{1}{i\hbar}\Delta V+2F_1\partial_q+F_2\partial_p\right\}
\mathcal{W}(q,p;t)=0,\label{Wignerequation}
}
where
\eq{
\Delta V= V\left(q+\frac{i \hbar}{2}\partial_p\right)-V\left(q-\frac{i \hbar}{2}\partial_p\right).
}
We write Eq. \eqref{Wignerequation} schematically as
\eq{
\left\{\partial_t+L_0+L_I+L_N(t)\right\}\mathcal{W}(q,p;t)=0,\label{Wignerequationshort}
}
where
\al{
L_0&=\frac{p}{m}\partial_q-m\omega^2q\partial_p\label{defL0}\\
L_I&=\frac{1}{i\hbar}\Delta V\label{defLI}\\
L_N(t)&=2F_1(t)\partial_q+F_2(t)\partial_p.
}
For specific potentials
\al{
V=\frac{\sigma}{3!}x^3 &\longrightarrow L_{I,1}=\frac{\sigma}{3!}(3q^2\partial_p -\frac14  \hbar^2 \partial_p^3)\\
V=\frac{\lambda}{4!}x^4 &\longrightarrow L_{I,2}=\frac{\lambda}{4!}(\hbar^2q\partial_p^3 -4 q^3 \partial_p).
}
We now strip off the time dependence in Eq. \eqref{Wignerequationshort} by defining
\eq{
\sigma(q,p;t)=e^{(t-t_0)(L_0+L_I)}\mathcal{W}(q,p;t),\label{definitionsigma}
}
which leads to the equation
\al{
\partial_t \sigma &= - e^{(t-t_0)(L_0+L_I)}L_N(t)e^{-(t-t_0)(L_0+L_I)}\sigma \\
&\equiv \Omega(t)\sigma
}
The formal solution of Eq. \eqref{Wignerequationshort} is
\al{
\nonumber \sigma(q,p;t)&=\mathbf{T}\left\{\exp\left[  \int_{t_0}^t dt_1\Omega(t_1)\right]\right\}\sigma(q,p;t_0)\\
&=\left(1+\int_{t_0}^t dt_1\Omega(t_1)+\int_{t_0}^t dt_1\int_{t_0}^{t_1} dt_2\Omega(t_1)\Omega(t_2)+...\right)
\sigma(q,p;t_0),\label{Wignersigmaequation}
}
where
\eq{
\sigma(q,p;t_0)=\mathcal{W}(q,p;t_0),
}
and $\mathbf{T}$ denotes the time ordering operation.
Now we want to average over the noise in Eq. \eqref{Wignersigmaequation}
by using the noise correlators \eqref{noisecorrelators}.  We assume that
the initial Wigner distribution is uncorrelated with the noise,
\eq{
\langle F_i(t) \mathcal{W}(q,p;t_0) \rangle = 0 = \langle F_i(t) \sigma(q,p;t_0) \rangle.
}
We then find that
\eq{
\langle \sigma(q,p;t) \rangle =\left\langle \mathbf{T}\left\{\exp\left[  \int_{t_0}^t dt_1\Omega(t_1)\right]\right\}\right\rangle \langle\sigma(q,p;t_0)\rangle.
}
The exponential can be expanded into powers of $\Omega(t)$. We use that
\al{
\langle \Omega(t)\rangle  &= 0\label{Omegacorriszero}\\
\nonumber \langle \Omega(t_1) \Omega(t_2) \rangle & = \delta(t_1-t_2) e^{(t_1-t_0)(L_0+L_I)} D^2(t_1) e^{-(t_1-t_0)(L_0+L_I)}
\label{Omega2correlator}.
}
where
\eq{
D^2(t_1)=4 f_{11}(t_1)\partial_q^2+2(f_{12}(t_1)+f_{21}(t_1))\partial_q\partial_p+f_{22}(t_1)\partial_p^2.
\label{derivativenoiseoperator}
}
Without loss of generality we now set $t_0=0$. Furthermore we use some shorthand notation where
\begin{align*}
\nonumber \int^t \Omega_1&\equiv\int_0^t dt_1 \Omega(t_1),\\
\nonumber \int^t\int^{t_1}\Omega_1\Omega_2&\equiv \int_{0}^t dt_1\int_{0}^{t_1} dt_2\Omega(t_1)\Omega(t_2),~\text{etc.}
\end{align*}
We therefore find that
\begin{align*}
\nonumber \langle\sigma(q,p;t)\rangle&=\left\langle\mathbf{T}\left\{\exp\left[  \int^t\Omega_1\right]\right\}\right\rangle\langle\sigma(q,p;0)\rangle\\
\nonumber &=\left(1+\int^t\int^{t_1}\langle\Omega_1\Omega_2\rangle+\int^t\int^{t_1}\int^{t_2}\int^{t_3}
\langle\Omega_1\Omega_2\Omega_3\Omega_4\rangle+...\right)\langle\sigma(q,p;0)\rangle\\
\nonumber &=\left(1+\int^t \int^{t_1}\langle\Omega_1\Omega_2\rangle+\int^t\int^{t_1}\int^{t_2}\int^{t_3} \langle\Omega_1\Omega_2\rangle\langle\Omega_3\Omega_4\rangle+...\right)\langle\sigma(q,p;0)\rangle\\
\nonumber &=\Biggl(1+\frac12 \int^te^{(L_0+L_I)t_1} D^2(t_1) e^{-(L_0+L_I)t_1}\\
\nonumber +\frac14\int^t\int^{t_1}&e^{(L_0+L_I)t_1} D^2(t_1) e^{-(L_0+L_I)t_1}e^{(L_0+L_I)t_2} D^2(t_2) e^{-(L_0+L_I)t_2}+...\Biggr)\langle
\sigma(q,p;0)\rangle.
\end{align*}
In the third line we have performed Wick contractions of the four- and higher-point correlators.
Notice that contractions other than the time ordered one vanish. The final line can be reexponentiated to the simple form
\begin{equation}
\langle\sigma(q,p;t)\rangle=\mathbf{T}\left\{\exp\left[  \frac12\int_{0}^t dt_1e^{(L_0+L_I)t_1} D^2(t_1) e^{-(L_0+L_I)t_1}\right]\right\}\langle\sigma(q,p;0)\rangle.\label{Wignersigmaequation}
\end{equation}
Differentiation of the last line gives
\begin{equation}
\partial_t \langle\sigma(q,p;t)\rangle=\frac12 e^{(L_0+L_I)t} D^2(t) e^{-(L_0+L_I)t}\langle\sigma(q,p;t)\rangle.\label{finalsigma}
\end{equation}
As $L_0$ and $L_I$ only depend on the classical quantities $q$ and $p$, they can be taken out of the noise average of $\sigma$.
This allows us to reexpress Eq. \eqref{finalsigma} in terms of the Wigner function $\mathcal{W}$ by using Eq. \eqref{definitionsigma},
\begin{equation}
\left\{\partial_t+L_0+L_I-\frac12D^2(t) \right\}\mathcal{W}(q,p;t)=0.\label{FPWignerequation}
\end{equation}
This is the generalised Fokker-Planck equation in phase space variables.
To remind the reader, $L_0$, $L_I$ and $D^2$ are defined in Eqs. \eqref{defL0}, \eqref{defLI} and
\eqref{derivativenoiseoperator}, respectively. The Wigner distribution can be used
to calculate field expectation values for operators of $x$ and $p$,
\eq{
\langle \mathcal{O}(x,p)\rangle=\int dq dp \mathcal{O}(q,p) \mathcal{W}(q,p;t).\label{Wignerexpectationvalue}
}
We emphasize here that the $x$ and $p$ on the lefthand side are operators,
whereas $q$ and $p$ on the righthand side are phase space variables.
One has to be careful when calculating phase space correlators
for which the ordering of $x$ and $p$ is important. The prescription is to move all
the $p$'s to the right in the operator $\mathcal{O}(x,p)$, and then apply Eq. \eqref{Wignerexpectationvalue}.
This is of no concern for our object of interest, the phase space area \eqref{phasespaceareaQM},
which does not contain any correlators involving the canonical commutator.\\
In principle one could make an ansatz for $\mathcal{W}(q,p;t)$ which solves
the Fokker-Planck equation \eqref{FPWignerequation}. In the case of a free particle
the solutions for the Gaussian Wigner distribution and related correlators
can be obtained analytically, see for example \cite{Koksma:2010zi} and \cite{Koksma:2010dt}.
For an interacting theory the non-Gaussian Wigner distribution cannot be solved exactly
and one has to resort to perturbative methods.

We note that the above Wigner formalism for a quantum mechanical particle
can be extended to quantum fields. This has been done by Habib \cite{Habib:1992ci}
for free fields. In principle the Wigner distribution $\mathcal{W}$
and the phase space area can be calculated numerically in a discretized version of
Eq. \eqref{FPWignerequation} for interacting quantum fields. However, we warn the reader
that it would be very challenging to find the Wigner function even numerically due
to the non-local nature of the interacting potential in Wigner space. Thus, although Eq.~\eqref{FPWignerequation}
provides an alternative method,
our proposed method in section \ref{sec:selfintpotential} of solving the stochastic Hamilton equations
(\ref{stochHamiltoneqphiINT}--\ref{stochHamiltoneqpiINT}) numerically provides a simpler way to
model cosmological decoherence.

\bibliography{DecoherenceStochasticInflation2}{}
\bibliographystyle{JHEP}

\end{document}